# Effect of Long-Range Dielectric Screening on Charge Transfer States

Michael J. Waters, Daniel Hashemi, and John Kieffer

**Abstract.**

Exciton dissociation at heterojunctions in photovoltaic devices is not completely understood despite being fundamentally necessary to generate electrical current. One of the fundamental issues for *ab initio* calculations is that hybrid interfaces combining materials with Wannier-Mott excitons and those with Frenkel excitons can easily require thousands of atoms to encompass the exciton-wave function. The problem is further exacerbated by a large permittivity difference at the interface, which requires meso-scale boundary conditions to accurately predict electrostatic potentials. For these reasons, we have constructed a model of excited states at hybrid interfaces based on an effective mass Schrödinger equation. In this continuum model, carrier wave functions are represented by their envelope function rather than resolving the atomic scale variations. Electrostatic interactions are accounted for using the Poisson equation. For our model system, we use a pentacene/silicon interface. Because carrier mobility is low in pentacene relative to silicon, the hole is frozen such that it only interacts with the electron though an immobile positive charge density. The inputs to this model are as follows: dielectric permittivities, electron effective masses, interfacial width, band alignment, and the hole wave function. We find that the energetic favorability of charge transfer states relative to bulk excitons is most easily controlled by band alignment. However, when both states have similar energies, interface proximity and electrostatics become important secondary means of tuning the relative stability of these states.

**Introduction.**

As one of our solutions to building a renewable energy infrastructure, photovoltaics (PV) have been the subject of intense research. While traditional silicon solar cells still dominate the commercial market, significant effort has been devoted to finding and perfecting novel solar cell designs that can surpass silicon in terms of price, performance, and lifespan. In terms of efficiency, classical semiconductors such as silicon, cadmium telluride, gallium arsenide, *etc.* remain unmatched,[1] but these materials have relatively costly and demanding production techniques. Many emerging PV technologies have aimed to reduce production costs by using easier to manufacture materials rather than improve performance as a means of becoming more economically viable. These emerging technologies such as perovskite, organic, quantum dot, and dye-sensitized solar cells, combine very different materials in heterojunctions. These heterojunctions pose particular theoretical challenges for predicting charge transport across them. The problem comprises both corrections to the energetics and transport theory. In our previous work[21] we discussed issues with the current state of the transport theory and addressed the issue of energetics with some semi-classical corrections. In this work, we have constructed a model based on the works of Stier *et al.* and Bolcatto *et al.* using the effective mass Schrödinger equation to self-consistently calculate excited states and energies at heterojunction interfaces.[2,3] To better understand the reasons for the theoretical difficulties in addressing heterojunction excited states, a review excited states in bulk materials is included.

In bulk materials, the two prototypical exciton types, Wannier-Mott and Frenkel, are characterized by opposing properties while both are still bound via the Coulomb interaction. Materials with high dielectric constants and low carrier effective masses typically have Wannier-Mott type excitons, which are characterized by being spread over many unit cells or molecules and having low binding energies that are quantized in similar way to the hydrogen atom. Materials with the opposite set of properties, low dielectric permittivity and high carrier effective masses, typically have Frenkel type excitons which are strongly bound and typically do not spread over more than a few atoms or molecules. From a theoretical standpoint, it is possible to calculate exciton binding energies for either type of material with good accuracy using the GW approximation and the Bethe-Salpeter equation. The difficulty in analyzing heterojunctions using electronic structure calculation methods is that describing the interface requires a large real-space calculation so as not to represent a periodic superlattice type structure. For a heterojunction using two Frenkel materials, it might be possible to use these methods because the Frenkel excitons are usually small enough to be computationally viable. But for a heterojunction containing a Wannier-Mott type material, the excited state volume is far too computationally demanding for most electronic structure methods. For heterojunctions containing two Wannier-Mott type materials, it is expected that excited state dissociation energies will be on the order of $k_B T$, like they are in the bulk, which makes their calculation less practically impactful for PV applications. This leaves the case of a hybrid heterojunction between a Wannier-Mott and Frenkel type material where the excited state dissociation energies are conceivably large enough to be of practical consideration and their real-space volumes large enough for them to be inaccessible to electronic structure calculations a fully atomic treatment. For these reasons, we implement an effective mass Schrödinger equation model based on Stier *et al.*[2] and the frozen hole approximation.[4]

Before continuing, it is important to address the semantics used in this work. When describing a bound excited state in relatively homogenous bulk material, we refer to the exciton binding energy as the energy required to separate the electron and hole in their lowest excited state from each other. However, in real devices, we are interested in the energy required to separate and collect carriers at opposite sides of the interface as this directly detracts from the open circuit voltage.[5] This leaves us with two different pictures: in the bulk the exciton must gain enough energy for the electron to reach the conduction band as shown in Figure 1(a). However at an interface, we want to know the energy required for the electron to reach the conduction band and then be collected far from the interface. Likewise, the hole needs to leave the interface to be collected in the opposite direction. For this reason we are reporting the dissociation energy for current collection across heterojunctions rather than the exciton binding energy. The difference in the LUMO and conduction band in Figure 1(b) is the band edge offset energy.

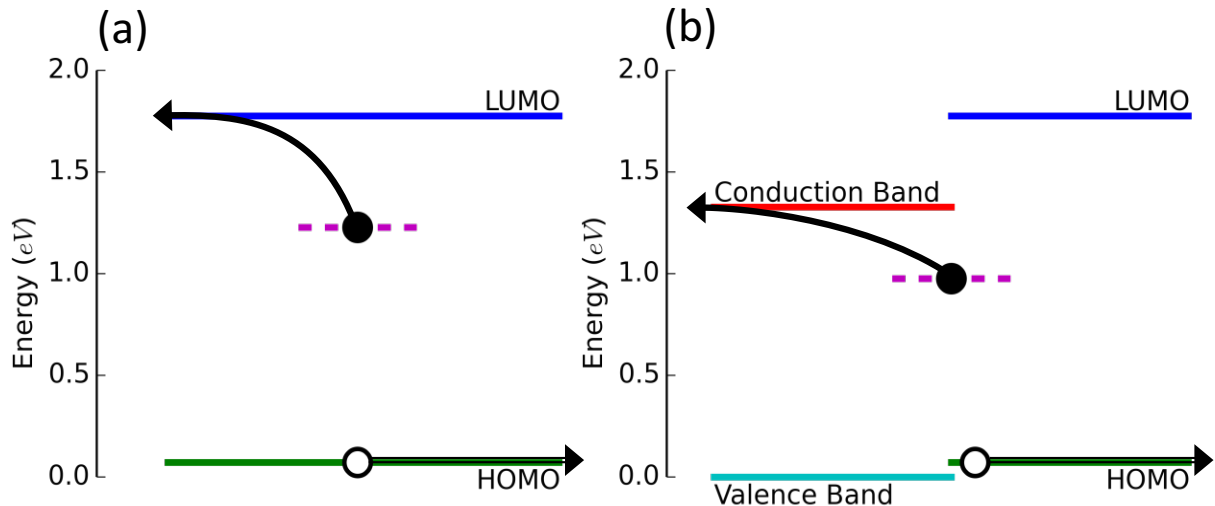

Figure 1. In (a), the dissociation energy of the excited state only requires that the electron and hole move far apart in any direction. In (b), the with the interface, the excited state dissociation energy for current collection is the energy required to move the electron into the conduction band and far away from the interface and the hole moving far in the opposite direction.

There is one final curious difference between Wannier-Mott and Frenkel excitons in their binding energy treatments. From a theoretical standpoint, the exciton binding energies of Wannier-Mott excitons are given as the lowest energy eigenvalue in a discrete spectrum of energy eigenvalues that varies as $E \propto {1}/{n^2}$. For Frenkel excitons, the binding energy is usually given as the Coulomb binding energy between the hole and the electron. The energy eigenvalues are, of course, a combination of kinetic energy and Coulomb interaction. The difference in formulation reflects that as Wannier-Mott excitons dissociate, their separate electron and hole wave functions are expected to become much more diffuse. Whereas for Frenkel excitons, their carriers' wave functions are expected to roughly retain their spatial extents as they move farther away from each other. In our hybrid system, we treat the dissociation energy with an electron kinetic energy that is assumed to approach zero as it travels far into the silicon, *i.e.*, the electron wave function is not confined and is behaves as it would in bulk silicon. We do not include the hole's kinetic energy because it is expected to not change significantly as it travels away from the interface since it is expected to remain confined to a few molecular orbitals.

**Theory.**

Our model development is not motivated by lack of accuracy of *ab initio* methods but rather by the desire to reduce computational cost. To simulate a Wannier exciton in a material where the excited stated extends over many unit cells, requires the inclusion of hundreds or thousands of atoms their electron wave functions, while we are only interested in the electron and hole wave functions. For these reasons we opted to use the effective mass Schrodinger equation approach,[2, 6, 7] which uses envelope wave functions as perturbations of the rapidly varying wave functions at the atomic scale, see Figure 2 below. This approach has been used with quantum dots[8] and junction tunneling.[6] Fundamentally, these models could be considered second principles models because they use first principles models for input parameters and trade exact electronic structure for more accurately capturing of the effect of meso-scale structures on electronic structure. In our implementation, we make no assumptions of wave functional forms. The Schrödinger equation is directly solved to obtain electron wave functions and energies. Materials properties are incorporated through our electron Hamiltonian.

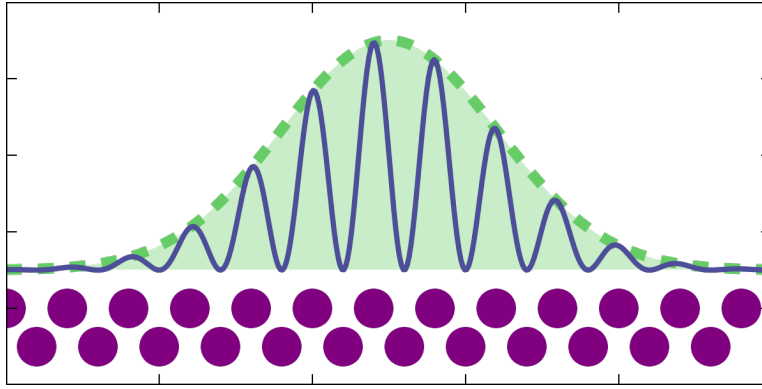

Figure. 2. Trading the rapidly varying wave function (filled blue) for the slowly varying one (dashed green line).

Our electron Hamiltonian is composed of four terms: kinetic energy, Coulomb interaction between the hole and the electron, the effective potential from the band edge offset, and the self-polarization due to the interface (also known as image potential or self-energy).

$$\widehat{H}_e = -\frac{\hbar^2}{2}\nabla \cdot \frac{1}{m_e^*(z)}\nabla + q_e\phi_h + V_{band} + \Sigma_e$$

(Equ. 1)

We include the effective mass inside the divergence operator because it produces better continuity across the interface due the change in effective mass.[6] The band edge offset potential, $V_{band}$, is a step function representing the difference between the LUMO of the pentacene and the conduction band of the silicon. The self-polarization potential, $\Sigma_e$, from the electron interacting with itself through the interface cannot be given in a closed form, which will be discussed in further detail in the methods section.

The potential field, $\phi_h$, created by the hole is obtained by solving the Poisson equation. By using a non-uniform the dielectric constant, we can capture the effect of the hole's 'image' charge on the electron in Equation 2. This formulation still requires the hole be represented by a charge density. From GW/BSE calculations; it is known that in pentacene crystals, excitons are delocalized over several intermolecular distances.[9]

$$q_h|\psi_h|^2 = -\nabla \cdot [\epsilon(z)\nabla\phi_h]$$

(Equ. 2)

It would be convenient to replace the charge density and the Coulomb term with a molecular pseudopotential that is tuned to give the correct exciton binding energy, however this would fail to capture the effect of the holes 'image' charge on the electron that we get from solving Poisson's equation. We examined several hole charge densities for their effect on the exciton binding energy and found that simply adapting a Gaussian profile results in an exciton binding energy that matches the first principles calculations of Sharifzadeh *et al.*[9]

**Algorithmic Implementation and Model Parameters.**

We implemented a self-consistent iteration scheme to solve the above Schrödinger equation in real space on a regular, finite-volume grid. While not a basis set in a traditional sense, we choose a discrete grid because standard atomic orbitals will not likely represent the effect of the abrupt transition near the interface. We also did not use plane waves because we are not using a periodic system. We implemented our model in FiPy[10] using PySparse's[11] precondition conjugate gradient solver (PCG) with Jacobi preconditioner as the backend solver and preconditioner, respectively, for solving both the Schrödinger equation and the Poisson equation. Higher energy states are calculated by performing numerically stable Gram-Schmidt orthonormalizations before each solver update of the wave function. This PCG solver is found to occasionally create non-converging oscillations in the wave function if too many iterations are performed between updates of the corresponding energy eigenvalue. However, increasing the number of iterations between eigenvalue updates yields a more quickly converging calculation. Our solution to improve the stability at a greater number of solver iterations per eigenvalue update is to mix old and new solutions for the wave function. Inspired by the DIIS method of Pulay,[12] we developed an efficient method for determining the optimal mixing parameter between old and new wave functions. We then coupled the optimal mixing parameter to the number of solver iterations per energy eigenvalue update via a rudimentary feedback algorithm (further described in Appendix B). With this implementation, minimum energy electron wave functions can be reliably calculated even if the wave function is assigned a random value at every grid point.

The classical result of the self-potential for a point charge near a sharp dielectric interface is inappropriate in this model due to the non-integrable singularity in the self-potential at the interface. We have elected to implement a diffuse dielectric interface model. To our knowledge, no closed form solution exists for the self-potential with a diffuse dielectric interface. Therefore, we implemented the numerical model for self-potential of a diffuse dielectric interface of Xue and Deng.[13] A plot of this dielectric constant and the image potential are shown in Figure 3. Our simplified implementation of the model of Xue and Deng is further detailed in Appendix A.

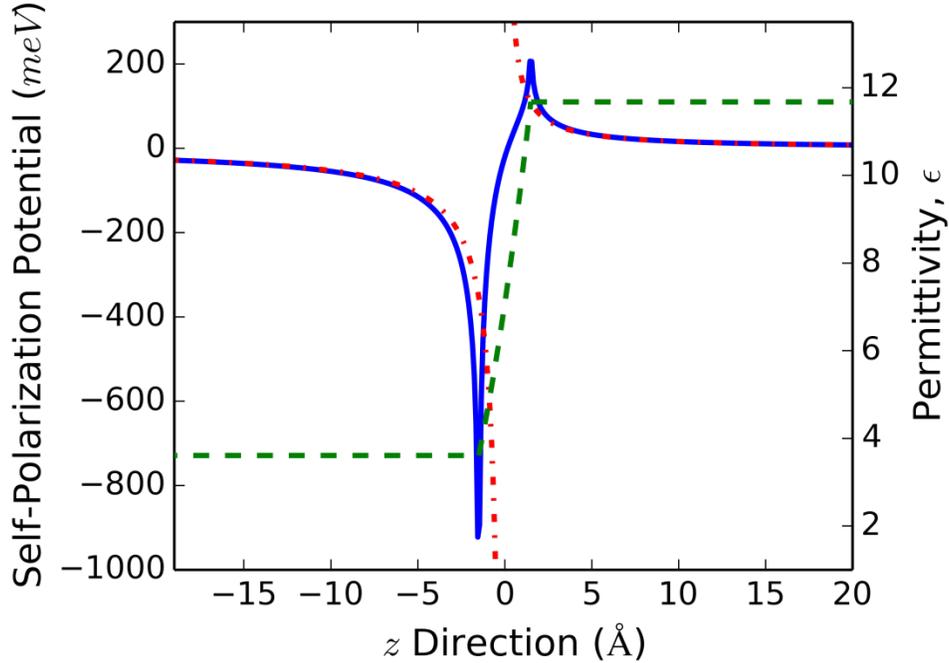

Figure 3. The dielectric permittivity relative to the interface is plotted with the green dashed line. The solid blue line is the calculated self-polarization energy and the dashed and doted red line is the divergent potential calculated for a sharp interface.

We created two versions of our simulation, a full three dimensional version and a two dimensional version, which, by assuming azimuthal symmetry about the axis normal to the interface, allows the azimuthal coordinate of a cylindrical coordinate system to be eliminated, and thus reduces the memory requirements. Both are shown in Figure 4. Both cases use, effectively, the same parameters. The interface between the pentacene and silicon lies at half the vertical height of the simulation size. For wave functions, the Dirichlet boundary condition of $\psi = 0$ was applied at the exterior faces. For the hole potential calculated from the Poisson equation, mixed boundary conditions are used, with the Dirichlet boundary condition applied at the top and bottom faces, and the Neumann boundary condition of zero electric field at the lateral exterior faces. This combination of boundary conditions on the electric potential allows the application of external potentials mimicking a real device. As with Stier *et al.*, we find that for most cases, the spatial requirement to capture the electron wave function is much smaller (≈ 10 x) than the spatial requirement to converge the hole potential. Using the cylindrical version, we find that exciton binding energies change by less than 1 meV for a radial and axial size of 800 Å. For the 3D version, this translates to 800 Å in the vertical direction and approximately 1600 Å in the lateral direction. Although, for the full 3D case, this size scale is prohibitive, due mostly to memory limitations. Smaller size scales proved useful in verifying that the ground state is indeed azimuthally symmetric and amenable to the less demanding 2D cylindrical model. For the grid spacing in the vertical direction, the sharpness of the self-polarization potential limits the spacing to 0.4 Å with an energy convergence of < 1 meV. In the radial direction, no such sharpness exists so the same energy convergence is achieved with a grid spacing of 1.0 Å. The Poisson equation is converged until PySparse reports a residual of less than $10^{-10}$. Wave functions are iterated until the relative change in energy eigenvalue is less than $10^{-5}$.

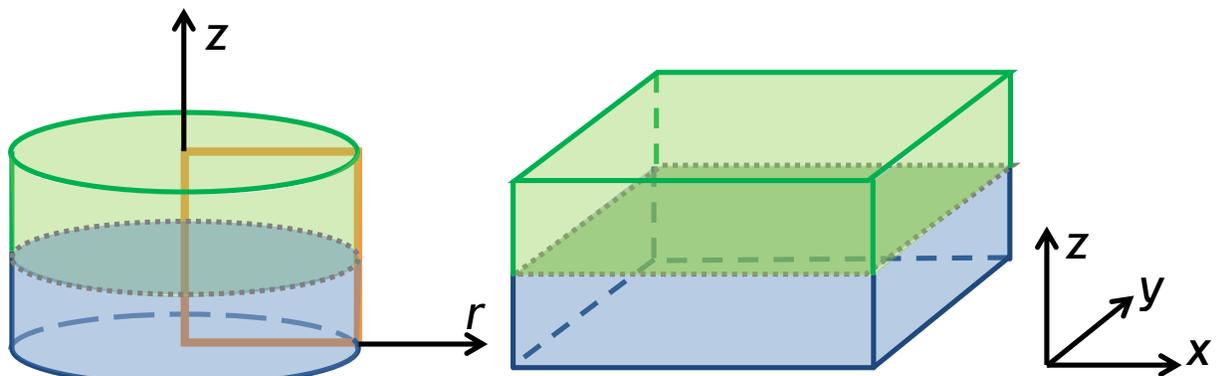

Figure 4. The two coordinate systems used in this work. For both, the green material represents the silicon and the blue represents the pentacene with the interface residing at half the z height. On the left, the azimuthally symmetric 2D cylindrical coordinates collapse the problem into a 2D grid inside the orange rectangle for which effectively larger simulations can be performed. On the right, the standard 3D Cartesian coordinates are pictured.

The materials parameters required for this model include the dielectric permittivities, the effective masses and the band offset between the LUMO of the pentacene and the conduction band of silicon. The electron effective masses in pentacene and silicon are anisotropic which, while implementable in our model, add unnecessary complexity to our model, and thus representative values for both are taken from Refs. 14 and 15. The static dielectric permittivity of pentacene is taken as the average of the two experimental values presented in Ref. 16. The LUMO of the pentacene molecule is found to be 478 meV above the conduction band of silicon in our previous work using the hybrid functional HSE06.[17] We chose a width, $\delta$, of 3 Å for the width of the dielectric interface because our previous work indicates that this width completely fits between a pentacene molecule and silicon surface. The spacing is also in agreement with the work of Cappellini and Del Sole which indicates that 3 Å is the size scale for the dielectric constant of silicon to develop bulk behavior.[18] These values are presented in Table 1 below:

Table 1. The six materials parameters used in the model.

| | |
|---|---|
| Static permittivity of silicon, $\epsilon_{Si}$ | 11.68 (Ref. 19) |
| Static permittivity of pentacene, $\epsilon_{pentacene}$ | 3.61 (Ref. 16) |
| Electron effective mass in silicon, $m^*_{Si}$ | 1.0 (Ref. 15) |
| Electron effective mass in silicon, $m^*_{pentacene}$ | 4.0 (Ref. 14) |
| LUMO pentacene / Si conduction band difference | 478 meV |
| Width of gradual dielectric interface, $\delta$ | 3 Å |

Before calculations of excited states at heterojunctions could be performed, the exciton binding energy of pentacene had to be reproduced. We use an azimuthally symmetric, Gaussian hole charge density because it can be used in both the two dimensional cylindrical and three dimensional models. The form of this charge density is described by Figure 5 and Equation 3 below:

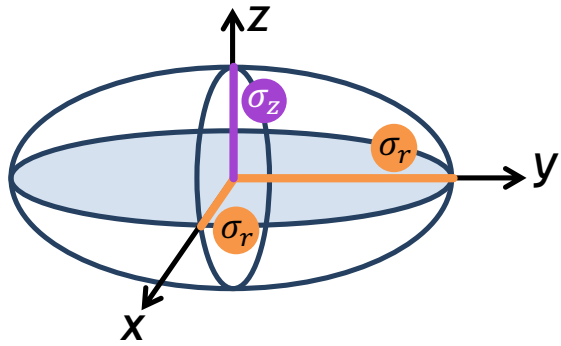

Figure 5. The azimuthally symmetric Gaussian density is roughly ellipsoidal where the principal axes in the x-y plane are the same.

$$\rho = \frac{q_h}{\sigma_z \sqrt{2\pi}} e^{-\frac{1}{2}\left(\frac{z-z_0}{\sigma_z}\right)^2} \frac{1}{2\pi\sigma_r^2} e^{-\frac{1}{2}\left(\frac{r}{\sigma_r}\right)^2}$$

(Equ. 3)

Exciton binding energy calculations are performed varying both the thickness $\sigma_z$ and width $\sigma_r$ of the hole charge density to find values that reproduce the magnitude of 0.5 eV known from GW/BSE calculations and experiment.[9] This is shown in Figure 6a. We opt for $\sigma_r = 6$ Å and $\sigma_z = 3$ Å because this corresponds approximately to twice the planar stacking distance of pentacene (≈ 3 Å) and twice the length (≈ 14 Å), and gives a Coulomb binding energy of 536 meV. This results in the electron wave function shown in Fig 6b.

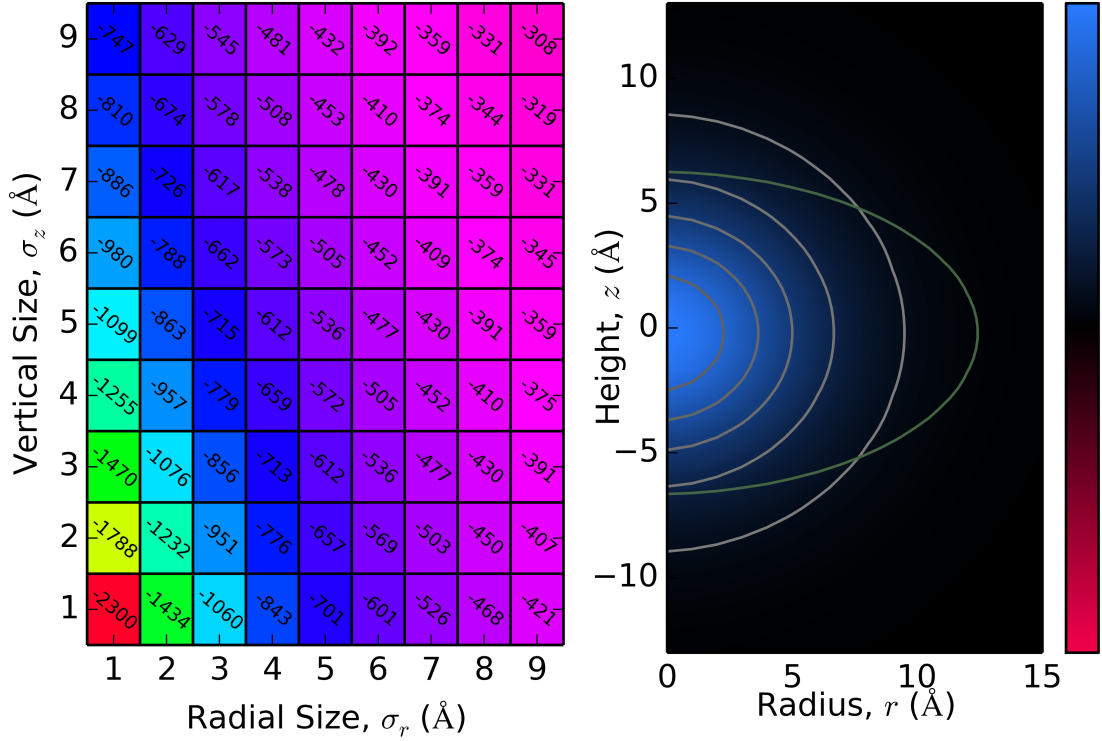

Figure. 6. (a) The exciton binding energy of pentacene versus frozen hole charge density thickness and width parameters. Energies are shown in the square at corresponding width and thickness. (b) The first electron state is shown on the right in blue/red for +/- values of the wave function. The green line is the isodensity contour containing 80% of hole charge density.

The first few electron wave functions of the Gaussian hole are calculated using a reduced 3D size of 200 Å x 200 Å x 100 Å. While the electrostatic energies are under-represented due to the poor Coulombic convergence from the smaller simulation size, the 3D radially symmetric electron wavefunctions calculated correspond to the ones calculated with the 2D cylindrical model. 3D states 1, 4, 5, and 10 match the 2D cylindrical wave functions states 1, 2, 3, and 4 in shape and are close in energies as shown in Table 2. The 3D states are consistently ~42 meV higher in energy eigenvalue. These results give confidence in our use of the 2D cylindrical model to find the exciton ground states.

Table. 2. The 3D electron wave functions and their respective energy eigenvalues for a hole in pentacene. The corresponding 2D wave functions and their eigenvalues are given.

| 3D State | | 3D Eigenvalue (meV) | Corresponding 2D State | | 2D Eigenvalue (meV) |
|---|---|---|---|---|---|
| 1 | 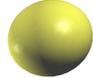 | -417 | 1 | 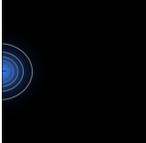 | -460 |
| 2 | 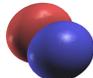 | -327 | - | - | - |
| 3 | 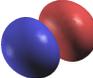 | -327 | - | - | - |
| 4 | 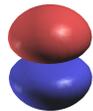 | -298 | 2 | 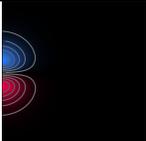 | -341 |
| 5 | 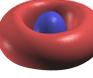 | -258 | 3 | 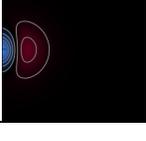 | -300 |
| 6 | 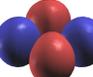 | -249 | - | - | - |
| 7 | 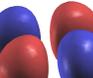 | -247 | - | - | - |
| 8 | 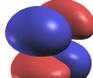 | -223 | - | - | - |
| 9 | 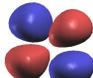 | -223 | - | - | - |
| 10 | 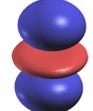 | -213 | 4 | 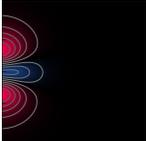 | -255 |

**Results and Discussion.**

The effect of the frozen hole position on the excited state dissociation energy has been examined by calculating the electron wave function at for various distances between the hole and the interface In Figure 7, we have plotted the exciton dissociation and Coulomb energies as a function of the distance from the interface with several of the electron wave functions shown. Moving the hole closer than 5 Å to the interface places a large fraction of the hole density in the silicon where it is not energetically preferable. The excited state is found to become more tightly bound as the hole moves closer to the interface. The electron wave function becomes more diffuse as the hole is moves farther away from the interface. This widening of the electron wavefunction as the hole moves farther away can only be due to the weakening Coulomb interaction in the silicon, since the Coulomb interaction is the only term in the Hamiltonian that depends on the hole position. We find that even if the hole is 100 Å from silicon, the electron wave function is still located in the silicon. In other words, the charge transfer state is always favorable over the exciton state. This result is surprising because the exciton binding energy is greater than the band offset energy. It seems that the band offset energy (pentacene LUMO/silicon conduction band difference) has a strong effect on the spatial distribution of the electron wave function.

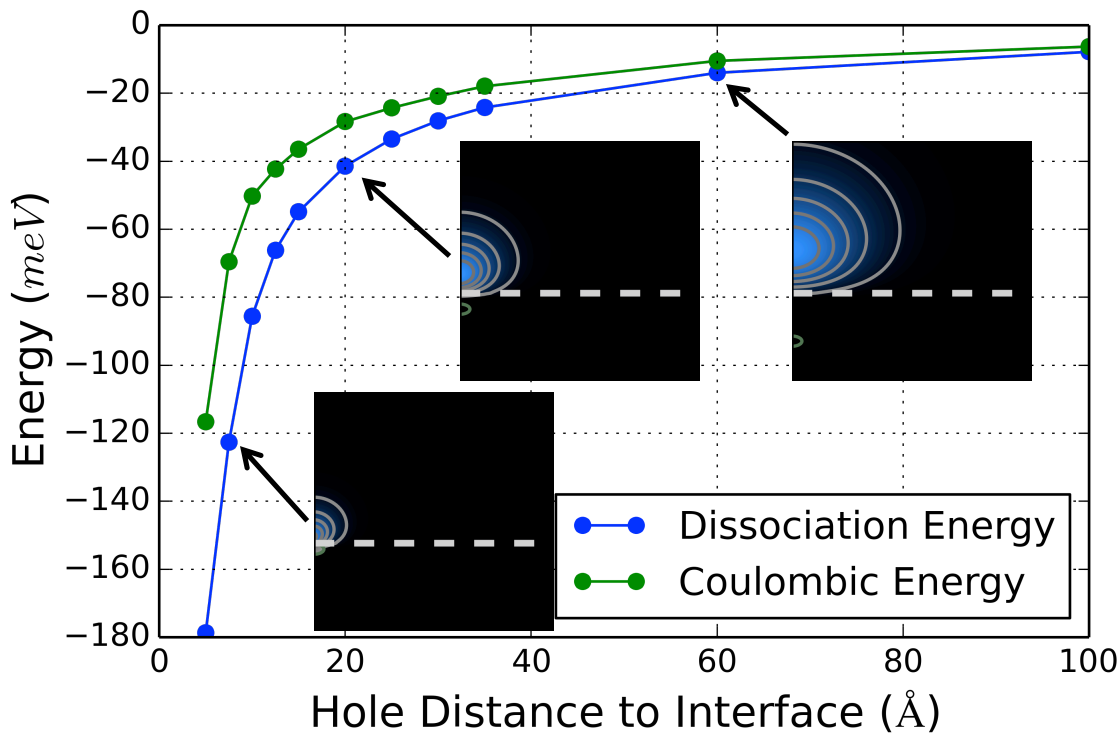

Figure 7. The dissociation energy (blue) and the Coulomb interaction (green) are plotted versus hole distance to the interface. Energies are reported such that negative values are binding. In the thumbnail images of the electron wave function (blue), the silicon is above the pentacene and the interface is denoted by the dashed white line. The small green half ellipse indicates the isodensity contour containing 80% of the hole. The same scale of 300 Å x 300 Å is used in each image.

To investigate the effect of band offset energy, a series of calculations were performed varying the band offset energy while the hole was fixed at 20 Å from the silicon. We find that as the band offset energy increases, the first state energy of the electron residing in pentacene rises linearly until a critical band offset energy is reached, where the electron switches into silicon and beyond which the electron energy remains constant, as shown in Figure 8. Note that Figure 8 shows actual data. Below the critical band offset energy, the electron distribution inside the pentacene remains unchanged since band offset energy is a uniform potential applied to the pentacene. Above the critical band offset energy, the electron distribution in the silicon changes little because the coulomb interaction remains constant and only the height of the potential barrier to entering the pentacene changes. In other words, because the band offset energy is a flat potential in pentacene and once the electron switches to the silicon the electron no longer feels the band offset energy as a flat potential but as an increasingly tall potential barrier. Curiously, the critical band offset energy required to cause the electron to reside in the silicon (~425 meV) is smaller than both the energy eigenvalue (460 meV) and the Coulomb energy (536 meV) of the first electron state in the bulk pentacene exciton.

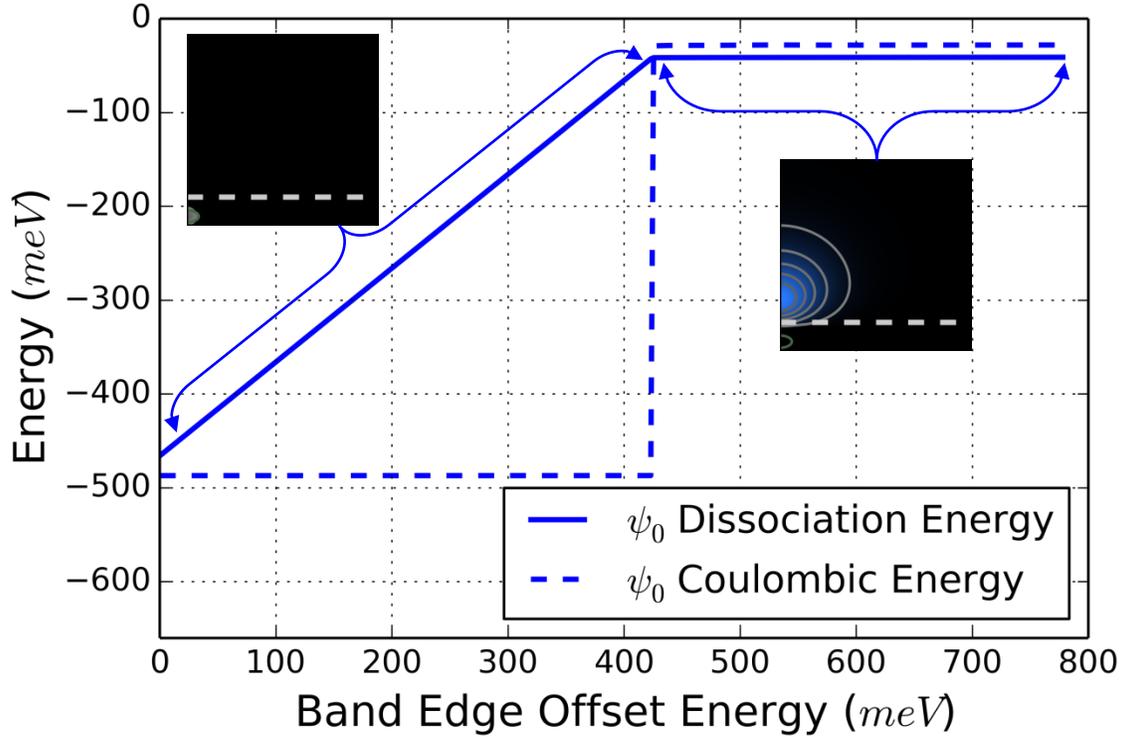

Figure. 8. The dissociation energy and coulombic energy are plotted versus the band edge offset energy. Below the critical value the electron sits in the pentacene and above, it resides in the silicon. In this sequence, the hole is fixed at 20 Å from the interface.

If the second electron energy level is calculated, it can be seen in Figure 9 that the second state's wave function below the critical band edge offset energy is essentially the same as the first state's wave function above the critical band edge offset energy. The first and second states are not degenerate states at the critical band edge offset energy because they cannot be truly linearly independent as there is always some non-zero wave function overlap. Above the critical band edge offset energy, the second wave function's similarity to the higher order electron wave functions seen in Table 2 above, suggests that non-azimuthally symmetric, lower energy dumbbell states probably exist.

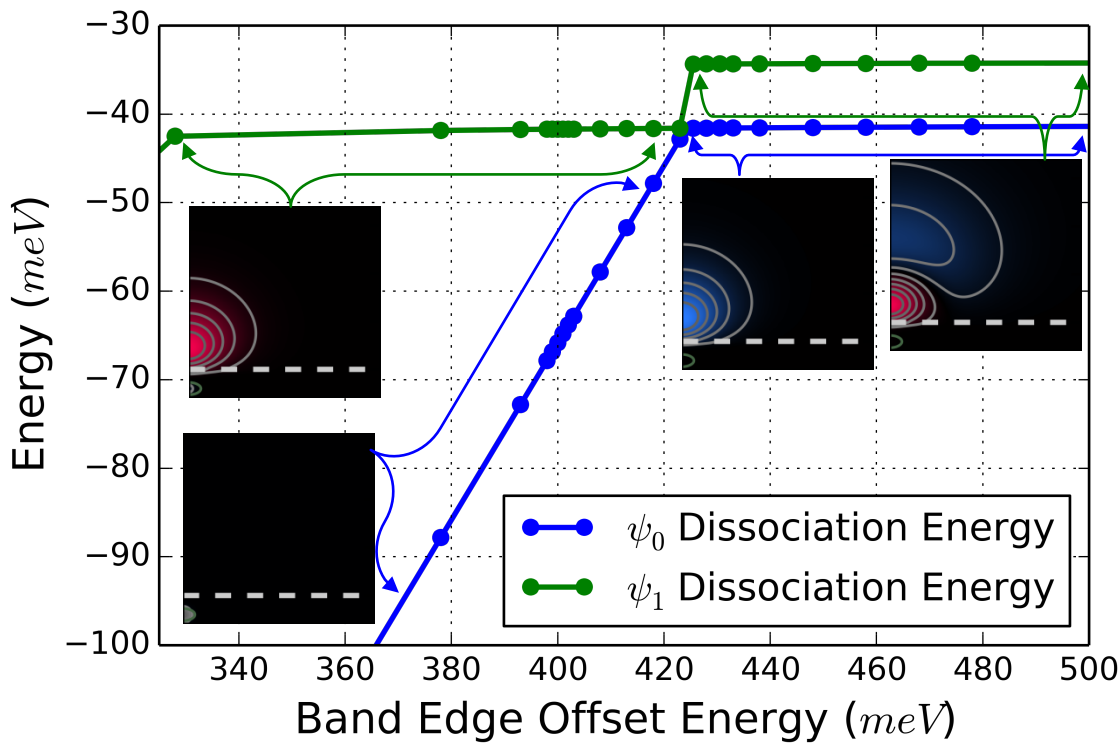

Figure 9. The dissociation energies and the wave functions of the first and second states of the electron are shown above as functions of the band edge offset energy. The dissociation energy of the first state is shown in blue and the second in green. Positive and negative values of the wave functions are shown in blue and red respectively. The critical band edge offset energy is located at ~ 425 meV.

The sharp transition between the electron residing in the pentacene and the silicon persists for all reasonable distances between the hole and interface. The critical band edge offset energy is calculated for a series of hole distances in Figure 10. This establishes a design parameter space for the energetically favorable separation of electron and hole. Beyond ~ 15Å, the critical band edge offset is only affected by the Coulomb interaction between the hole and electron. Below ~ 15Å, the electron wave function begins to occupy a hybrid state between what would be the two lowest states if the hole was further from interface. Importantly this establishes the boundary between stable excitons and stable charge transfer states in terms materials parameters.

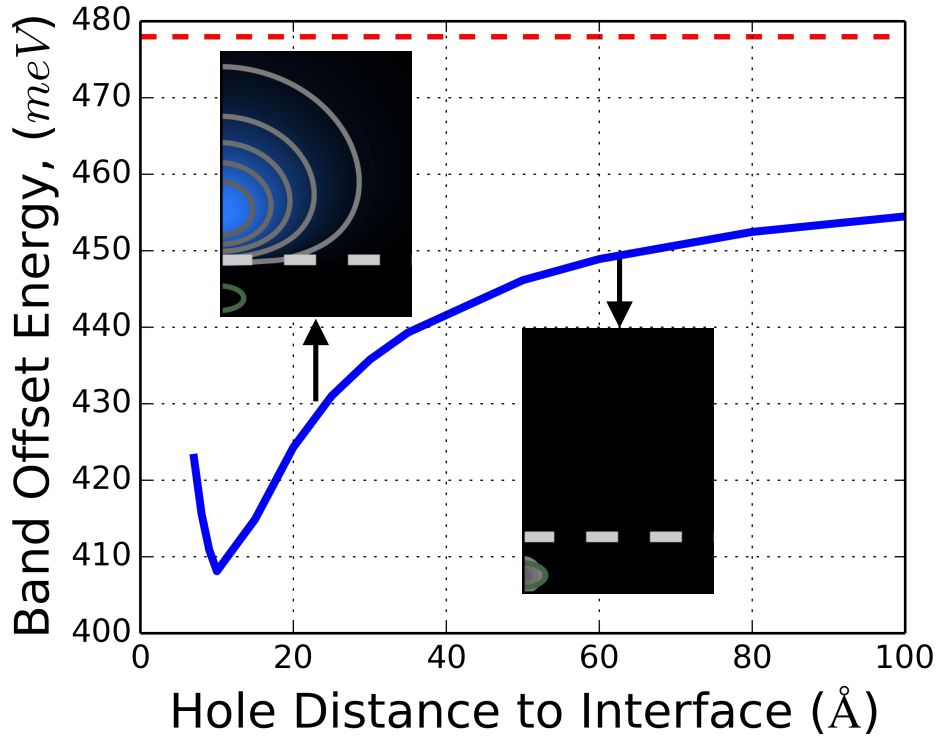

Figure 10. The critical band offset energy is plotted versus the hole distance to the interface. Above this line, it is favorable for the electron to reside in the silicon forming a charge transfer exciton. Likewise, below the line, the electron prefers to reside in the pentacene remaining a standard exciton. The dashed red indicated the band edge offset value used in Figure 7 which means that the charge transfer exciton is always favorable.

The model explored here is a significant step forward in capturing the long-range electrostatic interactions in excited states at interfaces for several reasons. The first is portability, this model can easily be used with other materials systems since only six materials and two hole parameters are required. The second is computational efficiency, most of these calculations were performed on a single workstation computer. The third is extensibility; while full size 3D calculations were found to be too expensive for this proof-of-concept implementation, off-the-shelf finite element codes with automatic mesh refinements exist which can alleviate this problem. Finally, there is no theoretical reason that prevents the addition of arbitrary potentials, using the time-dependent Schrödinger equation, or even coupling to phase field models of ferroelectrics. However, before other interface topologies can be implemented, an efficient computational means calculating self-polarization energies for arbitrary permittivity distributions must be found.

**Conclusions.**

We have created an efficient model for incorporating the effects of long-range electrostatic effects on excited states at interfaces. We find that above a critical band edge offset energy, the electron is more stable in the silicon and exhibits Wannier-Mott type behavior. Below this value, the electron prefers to remain in the pentacene and exhibits Frenkel type behavior. Interestingly, this critical value is less than either the bulk exciton binding energy or energy eigenvalue of the first electron state in the pentacene exciton. Above the critical band edge offset energy, the size of Wannier-Mott type electron wave function is proportional to the frozen hole distance to the interface. Because we establish stability regimes for the bulk-like exciton versus charge transfer states based on band offset and hole position, we offer a path for better control of interfacial transport of carriers. As far as we know, this is the first time this has been demonstrate in a hybrid inorganic/organic interface system.

**Appendix A.**

This numerical model for the self-energy of a point charge near a gradual dielectric interface is taken from Xue and Deng and has been reformulated to be concise and easier to implement.[13] The dielectric function is defined as a piecewise function:

$$\beta = \frac{\sqrt{\epsilon_{High}} - \sqrt{\epsilon_{Low}}}{\delta}$$

$$\alpha = \frac{\sqrt{\epsilon_{High}} + \sqrt{\epsilon_{Low}}}{2}$$

$$\epsilon(z) = \begin{cases} \epsilon_{Low}, & z \leq -\frac{\delta}{2} \\ (\alpha + \beta z)^2, & -\frac{\delta}{2} < z < \frac{\delta}{2} \\ \epsilon_{High}, & z \geq \frac{\delta}{2} \end{cases}$$

The self-polarization potential energy (or image potential energy) as a function of the source point charge position $z_s$ is given by:

$$\Sigma_e(z_s) = \frac{Q_s}{2} \int_0^\infty \frac{F(k, z_s)}{8\pi} dk$$

The charge of the self-interaction particle is $Q_s$ and the famous ½ term is due to this being a self-energy. The function $F$ is the difference between the Coulomb Green's function for the gradual interface and the ordinary Coulomb Green's in a Fourier-Bessel form. Unsurprisingly, $F$ is also piecewise with respect to the point charge position and is defined as follows.

$$F(z_s) = \begin{cases} F_{Low}, & z_s \leq -\frac{\delta}{2} \\ F_{Inter}, & -\frac{\delta}{2} < z_s < \frac{\delta}{2} \\ F_{High}, & z_s \geq \frac{\delta}{2} \end{cases}$$

$$F_{Low} = \frac{Q_s}{\epsilon_{Low} S(k)} e^{2kz_s} 2\beta \left[ k \left( \sqrt{\epsilon_{High}} e^{k\delta} - \sqrt{\epsilon_{Low}} e^{-k\delta} \right) - \beta \sinh k\delta \right]$$

$$F_{Inter} = \frac{Q_s}{\epsilon(z_s) S(k)} e^{-k\delta} 2\beta \left[ k \left( \sqrt{\epsilon_{High}} e^{-2kz_s} - \sqrt{\epsilon_{Low}} e^{2kz_s} \right) + \beta(e^{-k\delta} - \cosh 2kz_s) \right]$$

$$F_{High} = \frac{Q_s}{\epsilon_{High} S(k)} e^{-2kz_s} 2\beta \left[ k \left( \sqrt{\epsilon_{High}} e^{-k\delta} - \sqrt{\epsilon_{Low}} e^{\delta k} \right) - \beta \sinh k\delta \right]$$

We separate the dominator as a separate function since it has no dependence on $z_s$

$$S(k) = \left(-2k\sqrt{\epsilon_{High}} + \beta\right)\left(2k\sqrt{\epsilon_{Low}} + \beta\right) - \beta^2 e^{-2k\delta}$$

For our numerical evaluation of the self-polarization potential energy integral, we simply used Simpson's rule with $dk = 0.001$ Å$^{-1}$ and $k_{max} = 100$ Å$^{-1}$ which provides us with sufficient accuracy. For anyone wishing to reproduce this work, we suggest reading Refs. 13 and 20.

## Appendix B.

To improve convergence and stability when using self-consistent iteration to minimize the energy of an eigenstate wave function, we mix the old and the newly calculated wave functions using a linear mixing parameter, $\alpha$. Here we optimize this scheme then use the results to create a rudimentary solver iteration control algorithm which is very stable.

$$|\psi_{mixed}\rangle = \alpha|\psi_{new}\rangle + \beta|\psi_{old}\rangle$$
$$1 - \alpha = \beta$$

We have to then normalize the mixed wave function to get the final wave function:

$$|\psi_{final}\rangle = \frac{|\psi_{mixed}\rangle}{\sqrt{\langle\psi_{mixed}|\psi_{mixed}\rangle}}$$

What we really want is to minimize the final energy eigenvalue.

$$E_{final} = \langle\psi_{final}|\hat{H}|\psi_{final}\rangle$$

We can expand the final energy eigenvalue in terms of the mixing parameter:

$$E_{final} = \frac{\langle\psi_{mixed}|\hat{H}|\psi_{mixed}\rangle}{\langle\psi_{mixed}|\psi_{mixed}\rangle}$$

$$E_{final} = \frac{\alpha^2\langle\psi_{new}|\hat{H}|\psi_{new}\rangle + \beta^2\langle\psi_{old}|\hat{H}|\psi_{old}\rangle + \alpha\beta\left[\langle\psi_{new}|\hat{H}|\psi_{old}\rangle + \langle\psi_{old}|\hat{H}|\psi_{new}\rangle\right]}{\alpha^2\langle\psi_{new}|\psi_{new}\rangle + \beta^2\langle\psi_{old}|\psi_{old}\rangle + \alpha\beta[\langle\psi_{new}|\psi_{old}\rangle + \langle\psi_{old}|\psi_{new}\rangle]}$$

Since our old wave function is normalized and our new one can be easily normalized:

$$\langle\psi_{old}|\psi_{old}\rangle = 1$$
$$\langle\psi_{new}|\psi_{new}\rangle = 1$$

And since $\hat{H}$ doesn't change with iteration:

$$\langle\psi_{old}|\hat{H}|\psi_{old}\rangle = E_{old}$$
$$\langle\psi_{new}|\hat{H}|\psi_{new}\rangle = E_{new}$$

Also remember these identities:

$$\langle\psi_{new}|\psi_{old}\rangle^* = \langle\psi_{old}|\psi_{new}\rangle$$
$$\langle\psi_{old}|\hat{H}|\psi_{new}\rangle^* = \langle\psi_{new}|\hat{H}|\psi_{old}\rangle$$

We can clean up the final energy eigenvalue to a simpler form:

$$E_{final} = \frac{\alpha^2 E_{new} + \beta^2 E_{old} + \alpha\beta\left[\langle\psi_{old}|\hat{H}|\psi_{new}\rangle^* + \langle\psi_{old}|\hat{H}|\psi_{new}\rangle\right]}{\alpha^2 + \beta^2 + \alpha\beta[\langle\psi_{old}|\psi_{new}\rangle^* + \langle\psi_{old}|\psi_{new}\rangle]}$$

But for any complex number:

$$z^* + z = 2Re\{z\}$$

so:

$$E_{final} = \frac{\alpha^2 E_{new} + \beta^2 E_{old} + 2\alpha\beta Re\{\langle\psi_{old}|\hat{H}|\psi_{new}\rangle\}}{\alpha^2 + \beta^2 + 2\alpha\beta Re\{\langle\psi_{old}|\psi_{new}\rangle\}}$$

And for simplicity:

$$V_{on} = Re\{\langle\psi_{old}|\hat{H}|\psi_{new}\rangle\}$$
$$P_{on} = Re\{\langle\psi_{old}|\psi_{new}\rangle\}$$

So:

$$E_{final} = \frac{\alpha^2 E_{new} + \beta^2 E_{old} + 2\alpha\beta V_{on}}{\alpha^2 + \beta^2 + 2\alpha\beta P_{on}}$$

Also note that there are no complex terms here. We can substitute $\alpha$ for $\beta$:

$$E_{final} = \frac{\alpha^2 E_{new} + (1-\alpha)^2 E_{old} + 2\alpha(1-\alpha)V_{on}}{\alpha^2 + (1-\alpha)^2 + 2\alpha(1-\alpha)P_{on}}$$

We compute the derivative to minimize $E_{final}$ with respect to $\alpha$.

$$\frac{\partial E_{final}}{\partial \alpha} = \frac{2[(P_{on}-1)(E_{new}-E_{old})\alpha^2 + (2P_{on}E_{old} - 2V_{on} + E_{new} - E_{old})\alpha + (V_{on} - P_{on}E_{old})]}{[2\alpha^2(P_{on}-1) - 2\alpha(P_{on}-1) - 1]^2}$$

Since we are looking for extrema, we need the zeros of the numerator of this derivative since the denominator is always positive. Using the quadratic equation to find the zeros of the numerator gives two values of $\alpha$ that correspond to the extrema points.

$$(P_{on} - 1)(E_{new} - E_{old}) = a$$
$$2P_{on}E_{old} - 2V_{on} + E_{new} - E_{old} = b$$
$$V_{on} - P_{on}E_{old} = c$$

$$\alpha_{ex} = \frac{-b \pm \sqrt{b^2 - 4ac}}{2a}$$

We know which roots we want based on the concavity of the numerator, if the function is concave up, then the minima occurs at the larger root. If the function is concave down, the minima occurs at smaller root. We can write this simply:

$$\alpha_{optimal} = \frac{-b + sign(a)\sqrt{b^2 - 4ac}}{2a}$$

If the discriminant is negative there is no optimal value inside the bounds of (0,1]. If the computed $\alpha_{optimal}$, is less than 0, there is no optimal value inside the bounds of (0,1]. In both cases we simply defer to a preset minimum value of the mixing parameter. If the computed value of $\alpha_{optimal}$ is greater than one, we simply use $\alpha_{optimal} = 1$. We couple the value of $\alpha_{optimal}$ to the number of solver iterations used per step through a rudimentary control mechanism. If the minimum value of the mixing parameter we used, we decrease the number of solver iterations. If the value of $\alpha_{optimal}$ is greater than some cut-off value (usually 0.8) the number if solver iterations is increased. Using this method, the energies converge more quickly and less noisily while the solver rarely requires any hand tuning.